\documentclass[aps,prl,onecolumn,showpacs,groupedaddress,superscriptaddress,footinbib,longbibliography]{revtex4-2}

\usepackage{mathrsfs}
\usepackage{natbib,hyperref}
\setcitestyle{square,sort&compress,comma,numbers}
\hypersetup{colorlinks=true, citecolor=blue, urlcolor=blue, linkcolor=blue}
\usepackage[english]{babel}
\usepackage[utf8]{inputenc}
\usepackage{graphicx}
\usepackage[caption=false]{subfig}
\usepackage{amsmath}
\usepackage{amsfonts}
\usepackage{amssymb}
\usepackage{color,soul}
\setulcolor{red}
\usepackage{easyReview}
\usepackage{xcolor}
\usepackage{tikz}
\usetikzlibrary{calc}
\usepackage[percent]{overpic}
\usepackage[normalem]{ulem}
\renewcommand{\vec}[1]{\boldsymbol{#1}}

\begin{document}

\title{Disorder-induced persistent random motion and trapping of microswimmers}

\author{Mirko Residori}
\affiliation{Max Planck Institute for the Physics of Complex Systems, N\"othnitzer Stra{\ss}e 38,
01187 Dresden, Germany}
\author{Sebastian Aland}
\affiliation{Institute of Numerical Mathematics and Optimization, Technische Universit\"at Bergakademie Freiberg, Akademiestra{\ss}e 6,  09599 Freiberg, Germany}
\affiliation{Faculty of Informatics/Mathematics, HTW Dresden, Friedrich-List-Platz 1, 01069 Dresden, Germany}
\affiliation{Center for Systems Biology Dresden, Pfotenhauerstraße 108, 01307 Dresden, Germany}
\affiliation{Cluster of Excellence, Physics of Life, TU Dresden, Arnoldstraße 18, 01062 Dresden, Germany}
\author{Christina Kurzthaler}
\email{ckurzthaler@pks.mpg.de}
\affiliation{Max Planck Institute for the Physics of Complex Systems, N\"othnitzer Stra{\ss}e 38,
01187 Dresden, Germany}
\affiliation{Center for Systems Biology Dresden, Pfotenhauerstraße 108, 01307 Dresden, Germany}
\affiliation{Cluster of Excellence, Physics of Life, TU Dresden, Arnoldstraße 18, 01062 Dresden, Germany}

\begin{abstract} 
Microorganisms often move in confined, disordered environments, where hydrodynamic couplings can modify their transport behavior. Using extensive finite-element simulations, we investigate the dynamics of microswimmers -- modeled as squirmers -- in two-dimensional disordered porous media by resolving the full hydrodynamic interactions. We reveal that the deterministic coupling between activity, hydrodynamics, and disorder is sufficient to generate effective diffusive transport. Strong pushers and pullers become localized in the porous medium either by trapping at corners or dynamic trapping, depending on swimmer type and the obstacle packing fraction. Squirmers can escape from dynamic traps, leading to a prominent ``hopping-and--trapping'' dynamics. Strikingly, we find a pusher-puller asymmetry in the trapping probability that can be reversed by short-range swimmer–obstacle interactions, highlighting the sensitivity of transport to near-field effects. 
\end{abstract} 

\maketitle

Converting chemical energy into directed motion allows microorganisms to optimize their survival strategies and to perform biological functions~\cite{Tecon:2017, Suarez:2006, Guasto:2012, Dufrene:2020, Raina:2019}, while synthetic microrobots -- equipped with a self-propulsion mechanism -- are expected to play a major role for future nano-technological applications, ranging from targeted-drug-delivery~\cite{Erkoc:2019, Sedighi:2019} to bioremediation~\cite{Gao:2014, Adadevoh:2016}. At microscopic scales, the dynamics of these active agents are dominated by viscous forces~\cite{Lauga_2009}, imposing constraints on the swimming kinematics. Geometrical confinement of their surroundings further shape their transport behavior through hydrodynamic couplings and steric interactions~\cite{Elgeti_2015,Bechinger_2016, Spagnolie_2023}. Complex disordered media, omnipresent in natural and technological contexts, have been demonstrated to modify the motion of bacteria~\cite{Bhattacharjee_2019}, worms~\cite{Sinaasappel_2025},  and active colloids~\cite{Wu_2021} to intriguing hopping-and-trapping dynamics, while truncating their directional persistence and controlling large-scale bacterial transport through repeated surface-scattering events~\cite{Dehkharghani_2023}. Revealing the physical ingredients underlying these non-equilibrium processes not only provides fundamental insights into microbiological systems but also paves the way towards novel synthetic smart agents capable to adapt to their surrounding environment~\cite{Cichos:2020, Doostmohammadi:2023, Yang:2024, Baulin:2025, Loewen:2026}.

From a theory perspective, in the absence of hydrodynamic interactions transport of ``dry''  active particles through porous media has revealed localization in geometric pockets and corners driven by steric interactions~\cite{Chepizhko_2013, Bertrand_2018, Licata_2016, Volpe_2017, Kurzthaler_2021, Reichhardt_2014, Zeitz_2017}. 
Recent studies have demonstrated that self-propelled agents optimize their diffusion through porous media via tuning their reorientation mechanisms, such as run-and-tumble or run-reverse patterns, allowing them to truncate their trapping phases in geometric pockets and to hop through the pore space~\cite{Bertrand_2018, Licata_2016, Volpe_2017, Kurzthaler_2021, Reichhardt_2014, Mattingly_2025}. While this body of work highlights the importance of geometry, it typically neglects hydrodynamic couplings between a swimming agent and nearby surfaces, which could play an important role in confined, complex environments. 

Hydrodynamic interactions have been studied, both experimentally and theoretically, primarily for planar walls~\cite{Lauga_2006, Berke_2008, Spagnolie_2012, Contino_2015,Kantsler_2013, Lintuvuori_2016, Kuron_2019, Junot_2022} and isolated spherical obstacles~\cite{Takagi:2014, Spagnolie_2015}, which have been demonstrated to profoundly influence swimmer dynamics, leading to scattering, stable swimming states, and trapping near boundaries. These effects depend strongly on the swimmer's shape, its hydrodynamic (such as pusher/puller) signature, and tumbling behavior, and can be strongly amplified by the properties of the confinement, such as repulsive potentials~\cite{Lintuvuori_2016, Kuron_2019}, surface structure~\cite{Kurzthaler_2021_SM}, or boundary deformability~\cite{Garai_2025}. A paradigmatic model for describing microswimmers in this context is the \emph{squirmer} model, which was first introduced by Lighthill~\cite{Lighthill_1952} and later refined by Blake~\cite{Blake_1971a, Blake_1971b}. Within this framework, swimmer–surface interactions are known to depend sensitively on the prescribed surface-slip modes. In particular, it has been shown that hydrodynamic interactions alone can result in squirmer trapping at boundaries, alignment, or scattering from planar walls, as well as forward and backward orbiting around single obstacles~\cite{Lintuvuori_2016, Ahana_Thampi_2019, Thampi_2021, Li_2011, Nie_2023, Ishikawa_2004, Ishimoto_2013, Ishimoto_Crowdy_2017}. Extensions to periodic lattices have further demonstrated how geometric confinement guides swimmer trajectories and induces localization~\cite{Brown_2016, Chamolly_2017, Ishimoto_2023, Ramprasad_2025}. While these studies offer valuable insights into swimmer-surface interactions in controlled settings, microorganisms are typically found in much more complex environments, where disorder is an intrinsic property of the system. In such settings, swimmer–boundary interactions, geometric disorder, and activity are intricately coupled, making it difficult to anticipate the transport behavior of swimmers from single-obstacle studies. 

In this work, we isolate the effects of hydrodynamic interactions and spatial disorder on swimmer dynamics, while neglecting rotational noise or tumbling mechanisms of active agents. 
Combining a phase-field approach with extensive finite-element simulations, we study the motion of squirmers through a disordered porous medium. Our results reveal that deterministic hydrodynamic interactions with disordered obstacles give rise to multiple distinct trapping mechanisms, including static confinement, quasi-periodic orbital states, and geometric pocket trapping. Moreover, we find a pronounced asymmetry between pushers and pullers, leading to qualitatively different survival probabilities and transport regimes. Even in the absence of noise, swimmers exhibit diffusive, sub-diffusive, or super-diffusive motion, demonstrating that disorder-induced hydrodynamic reorientation alone is sufficient to generate anomalous transport and localization.

\section{The Model} 
\label{sec:model_and_method}
We consider a microswimmer navigating a disordered environment at low Reynolds number in two dimensions (2D). The swimmer is modeled as a disk-shaped squirmer of radius $a$, self-propelled by a prescribed tangential slip-velocity at its surface $\partial S$~\cite{Lighthill_1952, Blake_1971a, Blake_1971b}: 
\begin{align}
\vec{u}_S = B_1\left(1+\beta \vec{p}\cdot\vec{n}\right)\left(\vec{n}\vec{n}-\mathbb{I}\right)\cdot\vec{p}, \end{align} 
where $\vec{n}$ denotes the unit surface normal and $\vec{p}$ is the swimming direction. Further, $B_1$ represents the first squirming mode, which sets the swim speed in free space $U= B_1/2$~\cite{Molina_2013}. The swimmer type is characterized by the dimensionless squirming parameter $\beta$, measuring the stresslet strength, where $\beta < 0$ corresponds to pushers, $\beta=0$ to neutral swimmers, and $\beta > 0$ to pullers. The disordered medium is represented by $N$ randomly distributed non-overlapping disks of radius $R$~\cite{Torquato_2002} and characterized by the packing fraction $\phi= N\pi R^2/L^2$, where $L$ is the length of the domain. To prevent overlap between the squirmer and the obstacles we apply a short-ranged repulsive potential ensuring that the swimmer-obstacle distance exceeds the minimal distance $h_{\text{cut}}$. 

The spatially-varying fluid velocity and pressure fields, $\vec{u}(\vec{x},t)$ and $p(\vec{x},t)$, 
are governed by the quasi-steady incompressible Stokes equations:
\begin{align}
\mu\nabla^2\boldsymbol{u}=\nabla  p \quad {\rm and } \quad \nabla\cdot\vec{u}=0,
\end{align}
with viscosity $\mu$ and no-slip boundary conditions on the obstacle surfaces. The velocity on the squirmer surface ($S$) entails its translational and angular velocities, $\vec{U}$ and $\omega$, via: $\vec{u}= \vec{u}_S+\vec{U}+\omega \vec{e}_z \times \vec{r}$ (with $\vec{e}_z$ the unit normal to the 2D plane and $\vec{r}$ pointing from the swimmer center to its surface). The velocity and pressure fields and the velocities of the swimmer are obtained by solving the Stokes equations, subject to the force- and torque-free conditions on the swimmer:
\begin{align}
\int_{\partial S} \vec{n}\cdot\vec{\sigma}\ {\rm d}S =\vec{0}  \quad {\rm and } \quad   \int_{\partial S} \vec{r}\times \vec{n}\cdot\vec{\sigma}\ {\rm d}S=\vec{0},  \label{eq:torque-free}
\end{align}
where $\vec{\sigma}=-p\mathbb{I}+\mu(\vec{\nabla}\vec{u}+\vec{\nabla }\vec{u}^T)$ is the stress tensor. We employ a phase field approach for the squirmer and perform finite-element simulations with adaptive grid refinement~\cite{Alkaemper_2016, Vey_2007, Witkowski_2015, Engwer_2025, Dune, AMDiS}, taking into account periodic boundary conditions to model an effectively infinitely-large porous medium. 
The swimmer's instantaneous center of mass $\vec{x}_C(t)$
and orientation $\vartheta(t)$ are then updated according to the equations of motion: 
$\dot{\vec{x_C}}(t)=\vec{U}$ and $\dot{\vartheta}(t)=\omega$. Details of the numerical implementation and its validation with Refs.~\cite{Ishimoto_Crowdy_2017, Ahana_Thampi_2019} are provided in the Appendix.

\section{Results} 
\label{sec:results}
We perform extensive numerical simulations to investigate the dynamics of microswimmers, characterized by different squirming parameters $\beta=-8,\dots 8$. Specifically, we set $R=4a$ and $L=128a$, and select two cut-off  values $\delta = h_{\rm cut}/a = 1/20, 1/4$, which can lead to different squirmer-obstacle interactions~\cite{Kuron_2019}. To resemble  dilute and dense porous media, we  consider two packing fractions, $\phi = 0.15$ and $0.45$. 
We remark that from the perspective of a squirmer of radius $a$ interacting via a short-range repulsion of range $\delta$, the relevant excluded radius becomes $R_{\mathrm{eff}} = R + a + \delta$. 
The corresponding effective packing fraction, accounting for overlaps of the enlarged disks, can be extracted numerically and yields $\phi_{\mathrm{eff}}  \approx 0.69$
for the dense systems considered here.  Thus, although $\phi=0.45$ may appear moderately dilute, the swimmer effectively experiences a medium close to the percolation threshold~\cite{Torquato_2002}. We collect representative statistics by sampling $128$ trajectories, starting at different locations with different initial orientations, for four different porous media realizations and each parameter set. The initial position is restricted to lie within the percolating cluster of the medium.

\begin{figure}[tp]
    \centering
    \includegraphics[width=\linewidth]{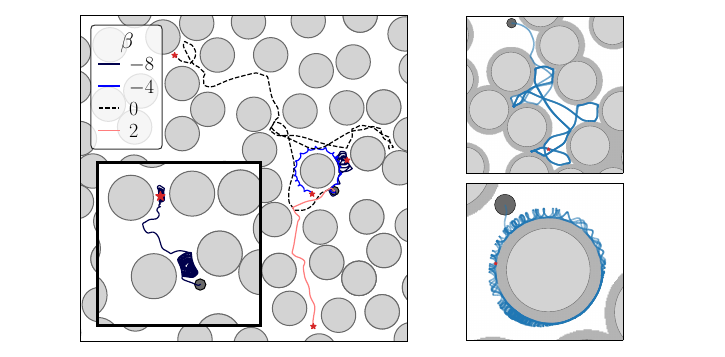}
    \caption{(\emph{Left}) Representative trajectories of squirmers in a disordered porous medium. Four swimmers with different squirming parameters $\beta$ start at the same position and orientation (dark gray disk). The end of the trajectory is marked by the red stars.  Here, we set the packing fraction to $\phi=0.45$ and use the cut-off $\delta = 1/20$. ({\it Inset}) A trajectory transitioning between dynamic trapping phases and free swimming motion. (\emph{Right}) Examples of dynamically trapped trajectories exhibiting quasi-periodic motion for $\beta=0$ (top) and $\beta=-4$ (bottom). In the former case, trapping is of topological origin, whereas in the latter it arises from hydrodynamic interactions. The shaded regions around the obstacles denote the effective disk radius, accounting for both the squirmer size and the short-range repulsive interaction.
    \label{fig:trajectories}
    }
\end{figure}

\subsection{Microswimmer dynamics: from swimming to trapping} To elucidate the impact of the squirming parameter on active motion through a dense disordered porous medium ($\phi=0.45$), we consider trajectories starting from the same initial position and orientation. Distinct swimming patterns emerge, as shown in Fig.~\ref{fig:trajectories} (\emph{left}). Neutral squirmers ($\beta=0$) undergo frequent scattering from obstacles, leading to long, curved trajectories through the pore space. Increasing $|\beta|>0$ results in trapping of squirmers in localized regions, whose nature depends on both the squirmer strength and the pusher/puller flow signature. For instance, a puller ($\beta=2$) is scattered by several obstacles before it ends up statically trapped at a fixed position due to geometric confinement. Differently, a pusher with $\beta=-4$ becomes trapped and orbits around a single obstacle. Decreasing the squirmer parameter ($\beta=-8$) leads to its localization between two obstacles, where it follows a quasi-periodic orbit, as the increase in squirming parameter enables hydrodynamic interactions with a second nearby obstacle. Here, it is important to clarify that a quasi-periodic orbit is referred to a trajectory that remains spatially confined and repeatedly follows a similar path, but does not close exactly on itself. Instead, the swimmer returns slightly displaced from its initial position after each cycle. For instance, in the single-obstacle case the orbital motion is accompanied by small oscillations, so successive revolutions do not exactly repeat the same trajectory. Similarly, between two obstacles the swimmer undergoes multiple scattering between them while following a nearly identical path, yet each cycle terminates at a slightly shifted location, see Fig.~\ref{fig:trajectories} (\emph{right}).

Inspection of our full dataset thus reveals two distinct states: exploration of the pore space and localization, either in the form of static trapping at obstacle boundaries or dynamic trapping in confined pockets. Transitions between the two states are also observed, see Fig.~\ref{fig:trajectories}({\it left-inset}). Here, the squirmer first becomes dynamically trapped and orbits between two obstacles. As the orbits are not closed, the swimmer eventually escapes and explores the medium. Subsequently, it enters a second dynamic trap formed by more closely spaced obstacles, where it remains for the remainder of the simulation, spending significantly longer times there than in the first trap.

These different behaviors are also observed for the dilute porous medium, yet with different statistical signatures depending on the squirmer parameter, the flow signature, and the short-range repulsive potential. 

\begin{figure}[tp]
    \includegraphics[width=0.65\textwidth]{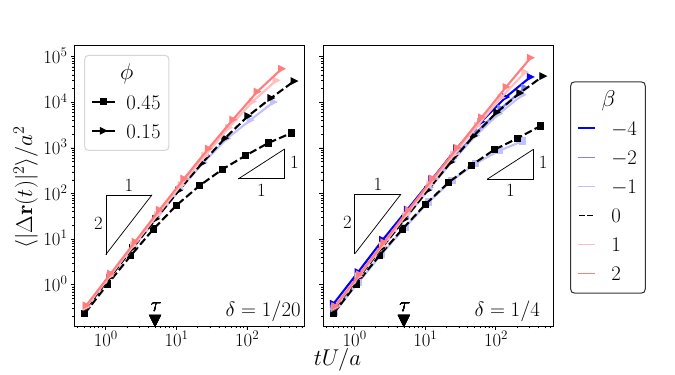}
    \caption{{\bf Exploring microswimmers.} Mean-squared displacement $\langle |\Delta\vec{r}(t)|^2\rangle/a^2$ for agents exploring the disordered medium. The left and right panel show results for $\delta = 1/20$ and $\delta=1/4$, respectively. Further, $\phi$ denotes the packing fraction and $\beta$ is the squirming parameter. The triangle marks the crossover time $\tau=(R+a)/U$, corresponding to the time it takes the agent to move the center-to-center contact distance.}
    \label{fig:msd_free}
\end{figure}

\subsection{Exploring agents: disorder-induced persistent random motion} To quantify the statistical properties of the different states, we first identify  parameter combinations ($\phi$, $\delta$, $\beta$) that allow squirmers to explore the environment without being trapped. We analyze the motion of these agents in terms of their mean-square displacement (MSD), $\langle |\Delta\vec{r}(t)|^2\rangle$, where $\Delta\vec{r}(t)=\vec{x}_C(t)-\vec{x}_C(0)$ is the displacement at time $t$.  Irrespective of the parameter set, our results show that at short times the MSDs obey $\langle |\Delta\vec{r}(t)|^2\rangle\approx U^2t^2$ due to their purely persistent swimming motion (Fig.~\ref{fig:msd_free}). Once the agents have traversed the center-to-center contact distance, $t\approx \tau:=(R+a)/U$, the MSDs start to deviate from a purely persistent behavior as hydrodynamic interactions with the randomly distributed obstacles impact their dynamics through modifying the swimming direction. Details of these interactions depend on the specific parameters.

In particular, for a small cutoff, $\delta=1/20$, and a dense porous medium, $\phi = 0.45$, free trajectories are observed only for neutral squirmers ($\beta=0$) (Fig.~\ref{fig:msd_free}, {\it left}). These display a diffusive behavior at long times $\langle |\Delta\vec{r}(t)|^2\rangle\sim t$, indicating that successive encounters with obstacles effectively randomize their swimming direction. For dilute systems ($\phi=0.15$), the MSD displays a super-diffusive regime at long times for $\beta\in\{-1, 1, 2\}$ with a clear trend: pullers  exhibit more pronounced super-diffusive behavior than pushers. This occurs because puller interactions with obstacles lead to fewer tangential sliding events and less frequent reorientations, in contrast to their pusher-counterpart, allowing them to preserve an effectively persistent motion for extended times.

By increasing the potential strength ($\delta=1/4$), we find that moderate pushers ($\beta\in\{-2, -4\}$) in dilute systems ($\phi=0.15$) also remain untrapped, as the repulsive potential effectively screens the strength of the hydrodynamic interactions (Fig. ~\ref{fig:msd_free}, {\it right}). The same qualitative trend is observed as for $\delta=1/20$: pullers exhibit more pronounced super-diffusive behavior than pushers,  with $\langle |\Delta\vec{r}(t)|^2\rangle\sim t^{2}$ for $\beta=2$, indicating a longer persistence of the swimming direction for pullers. In dense systems ($\phi=0.45$), neutral squirmers also display diffusive behavior while weak pushers exhibit slight sub-diffusion. This arises because hydrodynamic interactions with obstacles tend to align pushers tangentially to nearby surfaces. As a result, pushers spend longer times navigating close to obstacles, which reduces their effective displacement and slows their spatial exploration compared to a neutral squirmer.

We stress that the MSDs reported here reflect transient behaviors: due to numerical limitations, the asymptotic regime is not reached, and one should ultimately expect diffusive behavior in all cases at sufficiently long times.

\subsection{Localized agents: dynamic and static trapping}
\begin{figure*}[tp]
\centering
\begin{overpic}[width=\linewidth]{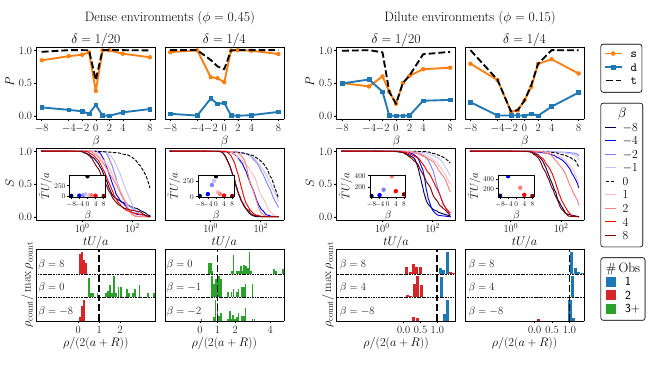}
 \put(-1, 47){(\bf{a})}
 \put(-1, 32){(\bf{b})}
 \put(-1, 16){(\bf{c})}
\end{overpic}
\caption{{\bf Localized microswimmers.} ({\bf a}) Trapping probability $P(\beta)$. Here, `s', `d', and `t' denote static, dynamic, and total trapping events. ({\bf b}) Survival probability $S(t)$ as a function of time $t$ for different squirming parameters~$\beta$. ({\it Inset}) shows the rescaled median of the trapping time $\tilde{T}$. ({\bf c})~Histogram of the normalized confinement radius $\rho/(2(a+R))$. 
({\bf a-c}) Different panels correspond to different packing fractions $\phi$ and repulsive potential strengths $\delta$.}
\label{fig:trapping}
\end{figure*}

Many microswimmers are unable to explore their environment for long times and instead become localized due to hydrodynamic coupling with obstacles and geometric confinement. Upon localization, we further characterize swimmer states through a confinement radius $\rho$ (i.e. the radius of gyration), computed from the spatial extent of each trajectory after its trapping time $t_{\mathrm{trap}}$(see Appendix). Swimmers with $\rho<a$ are classified as {\it statically trapped}, corresponding to localization at a fixed position, whereas swimmers with $\rho\geq a$ are classified as {\it dynamically trapped}, exhibiting persistent motion within a confined region. Using this criterion, we measure the trapping probability $P(\beta)$ as a function of the squirming parameter, shown in Fig.~\ref{fig:trapping} (a). To quantify the temporal aspects of this trapping process, we analyze the survival probability $S(t)$, defined as the probability that a swimmer is not localized in the porous medium up to time $t$.

We identify three generic features that are robust across packing fractions $\phi$ and cutoff distances $\delta$: (i)~Static trapping occurs more frequently than dynamic trapping. (ii)~Higher squirmer strength leads to faster trapping. (iii)~Neutral squirmers are, in almost all cases, the least likely to become trapped.

The simulations show that most neutral squirmers classified as statically trapped reach this state after swimming almost straight into an obstacle. In fact, this is essentially the only mechanism by which a neutral squirmer can be statically trapped: approaching a surface with an orthogonal orientation. Such a configuration is, however, unstable, since even a small perturbation in the swimming direction /would cause the squirmer to scatter away from the obstacle. In the numerical procedure, trajectories are terminated when both the translational and angular velocities fall below $\mathtt{tol}$, rather than reaching exactly zero. As a consequence, the measured static trapping probability for neutral squirmers is biased by the tolerance used to classify trajectories. Physically, this probability should be effectively zero. This bias is largely absent for nonzero values of $\beta$, because pushers and pullers undergo hydrodynamic reorientation near obstacles that increases their angular velocity, preventing them from satisfying the static trapping criterion.

\begin{figure*}[tp]
    \centering
    \includegraphics[width=\linewidth]{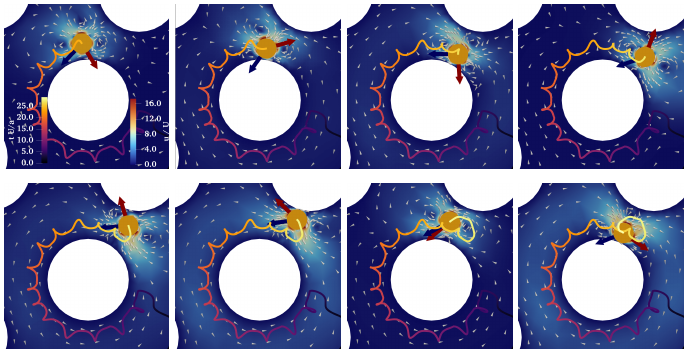}
    \caption{{\bf Velocity fields during a dynamic trapping event.} Snapshots of a strong pusher ($\beta=-8$) and small cutoff ($\delta_{\text{cut}} = a/20$) illustrating dynamic trapping between two obstacles. The blue arrow denotes the squirmer' s orientation $\vec{p}$, while the red arrow indicates the direction of its instantaneous velocity $\vec{U}$. The small arrows indicate the fluid velocity direction.
    }
    \label{fig:hydro}
\end{figure*}

\subsubsection{Trapping in dense environments}
In dense media ($\phi=0.45$), confinement is significantly enhanced and pullers are trapped with high probability for $\beta\neq 0$. Weak pushers, in contrast, can remain mobile provided the cutoff is sufficiently large ($\delta=1/4$), highlighting the screening effect of short-ranged repulsion (Fig.~\ref{fig:trapping}~{\bf a}). Consistently, survival statistics show that pushers persist longer than pullers at the same squirmer parameter. A higher cutoff markedly enhances survival, with a particularly strong effect for weak pushers, whose lifetimes become comparable to those of neutral squirmers (Fig.~\ref{fig:trapping}~{\bf b}). 

Analysis of dynamically trapped trajectories reveals two distinct localization mechanisms (Fig.~\ref{fig:trapping}~{\bf c}). For large squirmer parameters, trapping typically involves only two nearby obstacles and is characterized by small confinement radii, indicating hydrodynamically stabilized states in narrow gaps (Fig.~\ref{fig:trajectories}). Neutral squirmers instead display a qualitatively different behavior: their confinement radii are broadly distributed and trapping generally involves three or more obstacles, consistent with repeated scattering within the pore space; an example is displayed in Fig.~\ref{fig:trajectories} (\emph{right-top}).
Thus, while strong pushers and pullers are localized by hydrodynamic coupling to closely spaced obstacles, neutral swimmers and weak pushers become trapped only through purely geometric confinement.

\subsubsection{Trapping in dilute environments}
In dilute porous media ($\phi=0.15$), trapping depends sensitively on both the squirmer parameter $\beta$ and the short-range cutoff $\delta$ (Fig.~\ref{fig:trapping}~{\bf d}). For small cut-offs ($\delta=1/20$), strong pushers and pullers ($|\beta|\in\{4,8\}$) are almost certainly localized, but through distinct mechanisms: pushers predominantly undergo dynamic trapping in quasi-periodic orbital states, 
whereas pullers are mainly statically trapped with vanishing translational and angular velocities. Pullers that are dynamically trapped, are either stuck between two obstacles or orbit around one. For moderate squirmer strengths, trapping is primarily static and occurs more frequently for pushers than for pullers at the same squirmer magnitude. 

Increasing the cutoff to $\delta=1/4$ screens near-field hydrodynamic interactions and makes dynamic trapping between two obstacles rare. Strikingly, it also reverses the pusher--puller asymmetry: at fixed $|\beta|$, pullers become more likely localized than pushers. The trend is directly reflected in the median trapping times extracted from the survival probability (Fig.~\ref{fig:trapping}~{\bf b} ({\it inset})), which are larger for pullers than pushers at $\delta=1/20$ and larger for pushers than pullers at $\delta=1/4$. This behavior reflects the tendency of pullers to align orthogonally towards obstacle surfaces and enter near-hovering configurations, whereas pushers preferentially adopt tangential sliding states that facilitate escape for larger cut-off where hydrodynamic interactions are effectively weaker.

\subsubsection{Fluid dynamics of a dynamic trapping event}
To elucidate the microscopic origin of dynamic trapping, we examine a representative trajectory of a strong pusher confined between two obstacles (Fig.~\ref{fig:hydro}). For an isolated obstacle, a pusher swimming directly towards the surface cannot approach closer than a finite gap, where self-propulsion is balanced by hydrodynamic repulsion generated by the fluid pushed against the boundary. A slight tilt of the swimming direction destabilizes this head-on configuration, reorients the swimmer, and induces a lateral hydrodynamic attraction towards the obstacle, leading to a stable orbital state in which frontal repulsion and side-wise attraction balance. As a consequence of these hydrodynamic interactions, the squirmer orientation ($\vec{p}$, blue arrows) and its swimming velocity ($\vec{U}$, red arrows) can differ, see Fig.~\ref{fig:hydro}.

The resultant orbit is not strictly circular but exhibits persistent oscillations. When the swimmer approaches the surface too closely and reorients toward it, near-field hydrodynamic repulsion causes it to bounce away. Since the swimmer is not perfectly orthogonal to the surface, lateral attraction subsequently draws it back, resulting in a quasi-periodic oscillatory motion around the obstacle. During such an orbit, the swimmer may enter the hydrodynamic attraction range of a second nearby obstacle. In this case, while still oriented towards the first obstacle, the rear flow of the pusher generates a repulsive interaction with the second surface, causing the swimmer to slide along it in the opposite direction. Once sufficient separation is regained, the swimmer detaches and returns towards the first obstacle, thereby resulting in persistent confinement between the two obstacles, with alternating interactions with both boundaries.

\section{Summary and conclusions \label{sec:conclusions}} 
Combining a phase-field approach with extensive finite-element simulations, we have studied the dynamics of squirmers in dilute and dense disordered porous media, accounting for the full hydrodynamic interactions between the swimmer and the obstacles. We have revealed the emergence of persistent random motion of neutral squirmers due to hydrodynamically-induced reorientation by the nearby obstacles. We have further identified both static and dynamic trapping events of active agents, where microswimmers quasi-periodically trace certain paths within confined pockets for significant amounts of time. Examples range from single-obstacle orbiting to orbiting between two obstacles to moving in pockets confined by multiple obstacles. We find that these trapping characteristics strongly depend on the squirmer strength, the repulsive potential between the squirmer and the obstacles, and the packing fraction of the medium. Interestingly, our results show transient trapping phases, leading to hydrodynamically-induced ``hop-and-trap'' motion, reminiscent to `dry' active dynamics in porous media~\cite{Kurzthaler_2021}. 

Our approach, focused on disk-shaped microswimmers, can be readily extended to other swimmer shapes. This may allow us to understand the interplay of cell morphology and hydrodynamics for transport through porous media, where experiments have revealed that cell length and deformability can selectively enhance exploration or promote dead-end trapping depending on pore disorder~\cite{Sinaasappel_2025, Gao_2025}. It further provides a promising framework for unraveling why the motion of {\it E. coli} becomes rectified in ordered obstacle arrays, while spherical Janus colloids orbit around obstacles and stochastically switch to neighboring ones~\cite{Brown_2016}. 
Including tumbling mechanisms into our modeling approach, may allow us to elucidate the interplay of stochasticity in cell behavior and hydrodynamic couplings with the surrounding confinement~\cite{Junot_2022}. An additional open research endeavor entails the change in behaviors from two-dimensional systems, to quasi-two-dimensional channels, often used in microfluidic experiments, to three-dimensional dynamics.

While we have considered a quiescent fluidic environment, where flows are solely generated by the propulsion mechanism of active agents, many natural environments are subject to externally-imposed hydrodynamic flows~\cite{Wheeler:2019, Guasto:2012}. Controlled microfluidic set-ups of swimming microorganisms and theoretical studies of active agents in channel-geometries have revealed intriguing phenomena, such as rheotaxis (i.e. motion along/against a flow)~\cite{Tao:2026, Jing:2020, Mathijssen:2019, Peng:2020, gertack2026modes}, butterfly-like trajectories at channel constrictions~\cite{Waisbord:2021}, and accumulation of agents behind spherical obstacles~\cite{Mino:2018,Secchi:2020}. In complex environments, however, the resultant flow and vorticity fields are highly heterogeneous~\cite{Residori:2025}, leading to new effects. For example, these can induce curly swimmer trajectories~\cite{Creppy:2019, Das_2025},  trap point-like active agents via vorticity-induced reorientation in open channels~\cite{Das_2025}, or generate shear-induced Taylor dispersion~\cite{ AlonsoMatilla_2019}. Yet, the ramifications of the swimmer's own size and hydrodynamic signature on transport across porous channels in the presence of an externally-imposed flow remain largely unexplored. 

Beyond single-swimmer dynamics, our framework can be naturally generalized to account for hydrodynamic interactions between multiple squirmers. Collective effects are known to qualitatively modify transport properties in confinement, giving rise to clustering, enhanced diffusion, hydrodynamic bound states, and large-scale coherent motion~\cite{Marchetti_2013, Zoettl_2014,  Saintillan_2013}. In porous environments,
microswimmers could clog channels and thereby redirect flows, which could impact other active agents. Thus, the interplay of disordered porous media, hydrodynamic flows, and swimmer properties may produce exciting novel physical phenomena -- bridging clogging and self-organization -- which could have substantial implications for the formation of microbial communities and biofilms~\cite{Drescher:2013,Kurz:2022} and could inspire new tools for microswimmer sorting and filtration.  

\section{Acknowledgments} 
C.K. and M.R. acknowledge discussions with Sagnik Garai for theoretical insights and Simon Praetorious for the code development.
S.A. acknowledges support from DFG (grant 568855070).

\bibliography{references}

\section{Appendix}
The appendix contains a detailed description of the phase-field model,  its asymptotic analysis, and our numerical approach. We further show the validation of the model with known analytical results and provide details of the classification algorithm for the different trapping phases. 

\appendix
\renewcommand{\theequation}{A.\arabic{equation}}
\setcounter{equation}{0}

\subsection{Phase-field Model}
Let $S\subset\mathbb{R}^2$ denote the region occupied by the swimmer and $\Omega\setminus S$ the surrounding fluid domain. The governing equations read
\begin{subequations}
\begin{align}
\vec{\nabla}\cdot\vec{\sigma} &= \vec{0}, && \text{in } \Omega\setminus S, \label{eq:ge-momentum}\\
\vec{\nabla}\cdot \vec{u} &= 0, && \text{in } \Omega\setminus S, \\
\vec{u} &= \vec{u}_S+ \vec{U} + \omega\vec{e}_z\times\vec{r}, && \text{on } \partial S, \label{eq:velocity_bc}\\
\vec{u} &= \vec{0}, && \text{on }  \partial\Omega ,\label{eq:ge-momentum2}
\end{align}
\end{subequations}
where $\vec{u}$ and $p$ denote the velocity and pressure fields, respectively, $\vec{U}$ and $\omega$ are the translational and angular velocities of the swimmer, and $\vec{u}_S$ is a prescribed tangential slip velocity on the swimmer surface. The stress tensor is $\vec{\sigma} = -p\mathbb{I} + \mu(\nabla\vec{u} + \nabla\vec{u}^T)$ with viscosity $\mu$. We 
adopt a phase-field penalization formulation to avoid explicit tracking of the swimmer boundary, as required in immersed boundary approaches~\cite{aguillon_2012}. Our formulation follows the general framework of diffuse-domain and phase-field approaches, in which sharp boundaries are replaced by smooth indicator functions and interfacial conditions are imposed in a distributed manner; see Refs.~\cite{Li_2009, Aland_2010, Marth_2016,Reder_2022, mokbel2018phase}. In our model, the sharp interface $\partial S$ is replaced by a smooth indicator function $\psi$, and the equations are extended to the entire domain $\Omega$ in the following way:
\begin{subequations}
\begin{align}
\vec{\nabla}\cdot\vec{\sigma}
&=
\eta\,(\vec{U} + \omega\vec{e}_z\times\vec{r} + \vec{u}_S - \vec{u}),\label{eq:pf-momentum}
\\
\vec{\nabla}\cdot\vec{u} &= 0 ,
\\
\int_{\Omega}\eta\,(\vec{U} + \omega\vec{e}_z\times\vec{r} + \vec{u}_S - \vec{u}) \ \mathrm{d}\Omega&= \vec{0} ,\label{eq:pf-forcefree}
\\
\int_{\Omega}\eta\,\vec{r}\times(\vec{U} + \omega\vec{e}_z\times\vec{r} + \vec{u}_S - \vec{u}) \ \mathrm{d}\Omega &= \vec{0} ,\label{eq:pf-torquefree}
\end{align}
\end{subequations}
with the boundary condition $\vec{u}=\vec{0}$ on $\partial\Omega$.
The penalization function is defined as
\begin{align}
\eta = K\psi ,
\end{align}
where $K\gg1$ is a large constant. The indicator function is constructed from a smooth phase field
\begin{align}
\phi(\vec{x}) = \frac{1}{2}\left(1 + \tanh\left(\frac{a - \lVert \vec{r}\rVert}{\xi}\right)\right),\quad \psi(\vec{x}) = \lVert \nabla\phi\rVert = \frac{2}{\xi}\phi \,(1-\phi),
\end{align}
where $a$ is the swimmer radius and $\xi$ the interface thickness. In practice one chooses $K\sim\xi^{-2}$.

\subsection{Asymptotic Analysis}
In this section, we use matched asymptotic analysis to show that Eqs.~\eqref{eq:pf-momentum}-\eqref{eq:pf-torquefree} converge to the sharp-interface equations Eqs.~\eqref{eq:torque-free} and~\eqref{eq:ge-momentum}-\eqref{eq:ge-momentum2} as $\xi\rightarrow 0$. 
In this approach, we expand the variables in powers of the interface thickness $\xi$ in regions close to (inner expansion) and far (outer expansion) from the interface. The two expansions are matched in an intermediate region where both expansions are presumed to be valid (e.g., see \cite{caginalp1988dynamics,pego1989front} for a general description of the procedure). 

\textit{Outer expansion.} Let $S:=\{\vec{x}:\phi(\vec{x})\geq 0.5\}$. 
We assume that the stress $\vec{\sigma}$ has a regular expansion in $\xi$ in the fluid domain $\Omega\backslash S$, given by $\vec{\sigma}(\xi) = \vec{\sigma}^{(0)}+\xi \vec{\sigma}^{(1)} + \xi^2 \vec{\sigma}^{(2)} + ...$. 
Integrating Eq.~\eqref{eq:pf-momentum} over $\Omega$ and equating it with Eq.~\eqref{eq:pf-forcefree} gives
\begin{align}
	\int_\Omega \vec{\nabla}\cdot \vec{\sigma} \ \mathrm{d}\Omega= \vec{0}. 
\end{align}
Using the divergence theorem and inserting the expansion of $\vec{\sigma}$ gives to leading order
\begin{align}
	\int_{\partial\Omega} \vec{\sigma}^{(0)}\cdot \vec{n}\  {\rm d}S =  \vec{0}. \label{eq:asymptotic 1}
\end{align}

Moreover, away from the squirmer, we have $\phi=0$ to all orders in $\xi$, and so $\nabla\phi$ = 0 to all orders.
Therefore, the contribution of $\eta$ in Eq.~\eqref{eq:pf-momentum} becomes negligible outside the squirmer and the leading order of this equation yields $\nabla\cdot \vec{\sigma}^{(0)}=0$ in $\Omega\backslash S$. We therefore readily recover the sharp momentum balance equation \eqref{eq:ge-momentum}. 

Integrating this expression over $\Omega\backslash S$ and using the divergence theorem gives 
\begin{align}
	\int_{\partial \Omega \cup \partial S}  \vec{\sigma}^{(0)}\cdot \vec{n} \ {\rm d}S =  \vec{0}.
\end{align}
Subtracting Eq.~\eqref{eq:asymptotic 1}, we recover the sharp interface force balance \eqref{eq:torque-free}. Similarly, for the torque balance, we may take the cross product of Eq.~\eqref{eq:pf-momentum} with $\vec{r}$ and integrate over $\Omega$. Equating the result with Eq.~\eqref{eq:pf-torquefree} gives
\begin{align}
	\int_\Omega \vec{r}\times \nabla\cdot \vec{\sigma} = \vec{0}. 
\end{align}
Using the divergence theorem and the symmetry of $\vec{\sigma}$, and inserting its expansion gives to leading order
\begin{align}
	\int_{\partial\Omega} \vec{r}\times \vec{\sigma}^{(0)}\cdot \vec{n} \ {\rm d}S =  \vec{0}. \label{eq:asymptotic 2}
\end{align}
Moreover, away from the squirmer, a cross product of $\vec{r}$ with Eq.~\eqref{eq:pf-momentum} provides to highest order $\vec{r}\times \vec{\nabla}\cdot \vec{\sigma}^{(0)}=\vec{0}$ in $\Omega\backslash S$. Integrating over $\Omega\backslash S$ and using the divergence theorem gives 
\begin{align}
	\int_{\partial \Omega \cup \partial S}  \vec{r}\times \vec{\sigma}^{(0)}\cdot \vec{n} \ {\rm d}S =  \vec{0}.
\end{align}
Subtracting Eq.~\eqref{eq:asymptotic 2}, we recover the sharp interface torque balance \eqref{eq:torque-free}.

\textit{Inner expansion.}  It remains to be shown that the phase-field approach recovers the velocity boundary condition on the squirmer surface [Eq.~\eqref{eq:velocity_bc}].
Near this boundary, we introduce a local coordinate system 
\begin{align}
\vec{x}(\vec{s}, z;\xi)=\vec{X}(\vec{s})+\xi z \vec{n}(\vec{s})
\end{align}
where $\vec{X}(\vec{s})$ is a parametrization of the interface, $z=r(\vec{x})/\xi$ is the stretched variable, and $r$ is the signed distance from the point $\vec{x}$ to $\partial S$, which is taken to be positive inside the squirmer. We then assume that all variables can be written as functions of $z$ and $\vec{s}$ and that in these coordinates the variables have regular expansions in $\xi$. That is, for the velocity field
\begin{align}
\hat{\vec{u}}(z, \vec{s} ; \xi) \equiv \vec{u}(\vec{x} ; \xi)=\vec{u}(\vec{X}(\vec{s})+\xi z \vec{n}(\vec{s}) ; \xi)
\end{align}
and the inner expansion is
\begin{align}
\hat{\vec{u}}(z, \vec{s} ; \xi)=\hat{\vec{u}}^{(0)}(z, \vec{s})+\xi \hat{\vec{u}}^{(1)}(z, \vec{s})+\xi^2 \hat{\vec{u}}^{(2)}(z, \vec{s})+\cdots
\end{align}
The nabla operator in the local coordinate system becomes 
$$
\vec{\nabla} = \frac{1}{\xi}\vec{n}\partial_z + \vec{\nabla}_\Gamma,
$$
where $\vec{\nabla}_\Gamma$ is the covariant surface derivative. 

Inserting these local derivatives, we can consider the inner expansions of equation \eqref{eq:pf-momentum}.
Assuming that the penalization constant $K$ scales as $\xi^{-2}$,
we find at highest order $1/\xi^3$: 
\begin{align} 
\partial_z \hat{\phi}^{(0)} \left(\hat{\vec{U}}^{(0)}+\hat{\omega}^{(0)}\vec{e}_z\times \hat{\vec{r}}+ \hat{\vec{u}}_S - \hat{\vec{u}}^{(0)} \right) = \vec{0}. 
\end{align}
Since $\partial_z \hat{\phi}^{(0)} > 0 $ for all $z$, we recover Eq.~\eqref{eq:velocity_bc}: 
\begin{align}
\hat{\vec{U}}^{(0)}+\
\hat{\omega}^{(0)}\vec{e}_z\times \hat{\vec{r}}+ \hat{\vec{u}}_S = \hat{\vec{u}}^{(0)} . 
\end{align}
We may further take the derivative of this expression with respect to $z$ to obtain to highest order
\begin{align}
\partial_z \hat{\vec{u}}_S = \partial_z \hat{\vec{u}}^{(0)},
\end{align}
where we have used that $\partial_z \hat{\vec{U}}^{(0)}=\vec{0}$, $\partial_z \hat{\omega}^{(0)}=0$, $\partial_z \hat{\vec{r}} = \hat{\vec{n}}\xi$. 
Matching inner and outer solution in an overlap region where both expansions are valid, leads to the typical matching condition:  $\vec{u}$ is continuous across the surface if $\partial_z \hat{\vec{u}}^{(0)}=0$  (see \cite{pego1989front}).
Hence, if the imposed slip velocity $\vec{u}_S$ is continuous across the squirmer surface, then we have global continuity of the resulting velocity field $\vec{u}$. 

\subsection{Numerical approach}

\begin{figure*}[tp]
    \centering
    \includegraphics[width=0.76\textwidth]{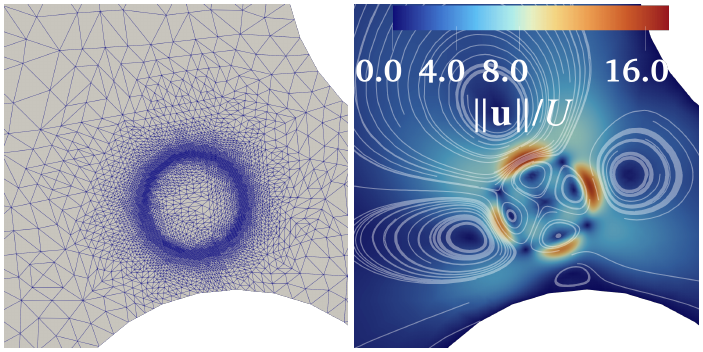}
    \caption{
 \emph{Left:} Adaptive mesh refinement used to resolve steep gradients and hydrodynamic interactions in the phase-field formulation.
 \emph{Right:} Flow field generated by a pusher ($\beta=-8$) in a confined geometry, rescaled by the free-space swimming speed.
}
    \label{fig:numerics}
\end{figure*}

The phase-field equations are discretized using the finite-element method. 
The diffuse interface has thickness $\xi$ and introduces steep gradients in the indicator function $\psi$ and the penalization field $\eta$. Accurate resolution of this region requires locally refined meshes. Finite elements allow straightforward adaptive mesh refinement, enabling the interface region to be resolved while maintaining a moderate number of degrees of freedom in the bulk fluid, see Fig.~\ref{fig:numerics}.

We next derive the weak formulation and the corresponding discrete system.
Let $H_0^1(\Omega)$ denote the usual Sobolev space. In the weak formulation we aim at finding $(\vec{u},p,\vec{U},\omega)\in H_0^1(\Omega)\times L^2(\Omega)\times\mathbb{R}^2\times\mathbb{R}$ such that
\begin{subequations}
\begin{align}
\int_\Omega \left(\mu\nabla\vec{u}:\nabla\vec{v} - p\,\nabla\cdot\vec{v} -
\eta\,(\vec{U}+\omega\vec{r}^{\perp}-\vec{u})\cdot\vec{v}\right) \ \mathrm{d}\Omega
&=
\int_\Omega \eta\, \vec{u}_S\cdot\vec{v} \ \mathrm{d}\Omega,
\\
-\int_\Omega q\nabla\cdot\vec{u} \ \mathrm{d}\Omega &= 0,
\\
-\int_\Omega \eta\,(\vec{U}+\omega\vec{r}^{\perp}-\vec{u})\cdot\hat{\vec{e}}_i\ \mathrm{d}\Omega
&=
\int_\Omega \eta\,\vec{u}_S\cdot\hat{\vec{e}}_i\ \mathrm{d}\Omega,\quad i=\{1,2\}
\\
-\int_\Omega
\eta\,\vec{r}^{\perp}\cdot(\vec{U}+\omega\vec{r}^{\perp}-\vec{u})\ \mathrm{d}\Omega
&=
\int_\Omega \eta\,\vec{r}^{\perp}\cdot\vec{u}_S \ \mathrm{d}\Omega,
\end{align}
\end{subequations}
for all test functions $(\vec{v},q)$. Here, we denote $\vec{r}^\perp = \vec{e}_z\times \vec{r} = (-r_2, r_1, 0)$.

Let $\{\boldsymbol{N}_j\}_{j=1}^{N_u}$ denote the velocity basis functions and
$\{Q_i\}_{i=1}^{N_p}$ the pressure basis functions. The discrete velocity and pressure fields are
\begin{align}
\vec{u}_h = \sum_{j=1}^{N_u} u_j \boldsymbol{N}_j
\quad {\rm and} \quad 
p_h = \sum_{i=1}^{N_p} p_i Q_i .
\end{align}

The resulting algebraic system takes the block form
\begin{align}
\begin{bmatrix}
\vec{A} & \vec{B}^T & \vec{C}^T \\
\vec{B} & \vec{0} & \vec{0} \\
\vec{C} & \vec{0} & \vec{M}
\end{bmatrix}
\begin{bmatrix}
\vec{u}_h \\
p_h \\
\vec{s}
\end{bmatrix}
=
\begin{bmatrix}
\vec{f}_h \\
0 \\
\vec{t}
\end{bmatrix},
\end{align}
where $\vec{s} = (U_1,U_2,\omega)^T$. The matrix entries are defined as
\begin{subequations}
\begin{align}
A_{ij} &= 
\int_\Omega  \left( 
\nu \vec{\nabla} \boldsymbol{N}_i : \vec{\nabla} \boldsymbol{N}_j
+
\eta\, \boldsymbol{N}_i \cdot \boldsymbol{N}_j\right) \ \mathrm{d}\Omega, \\
B_{ij} &= 
-\int_\Omega 
Q_i\, \vec{\nabla}\cdot \boldsymbol{N}_j \ \mathrm{d}\Omega.
\end{align}
\end{subequations}
The rigid-body coupling matrix reads
\begin{align}
C =
\begin{bmatrix}
-\int_\Omega \eta\, \boldsymbol{N}_j \cdot \hat{\vec{e}}_1 \ \mathrm{d}\Omega\\
-\int_\Omega \eta\, \boldsymbol{N}_j \cdot \hat{\vec{e}}_2 \ \mathrm{d}\Omega\\
-\int_\Omega \eta\, \vec{r}^{\perp} \cdot \boldsymbol{N}_j \ \mathrm{d}\Omega
\end{bmatrix},
\end{align}
where $\hat{\vec{e}}_1$ and $\hat{\vec{e}}_2$ are the Cartesian basis vectors.
The matrix $M$ is given by
\begin{align}
M =
\begin{bmatrix}
\int_\Omega \eta \ \mathrm{d}\Omega & 0 & -\int_\Omega \eta\, r_2 \ \mathrm{d}\Omega\\
0 & \int_\Omega \eta \ \mathrm{d}\Omega& \int_\Omega \eta\, r_1 \ \mathrm{d}\Omega\\
-\int_\Omega \eta\, r_2\ \mathrm{d}\Omega & \int_\Omega \eta\, r_1 \ \mathrm{d}\Omega& \int_\Omega \eta \|\vec r\|^2\ \mathrm{d}\Omega
\end{bmatrix},
\end{align}
and the right-hand side vectors are
\begin{align}
f_{h,j} = \int_\Omega \eta\, \vec{u}_S \cdot \boldsymbol{N}_j \ \mathrm{d}\Omega\quad {\rm and} \quad 
\vec{t} =
\begin{bmatrix}
\int_\Omega \eta\,  \vec{u}_S \cdot \hat{\vec{e}}_1\ \mathrm{d}\Omega \\
\int_\Omega \eta\, \vec{u}_S \cdot \hat{\vec{e}}_2\ \mathrm{d}\Omega \\
\int_\Omega \eta\, \vec{r}^{\perp}\cdot \vec{u}_S\ \mathrm{d}\Omega
\end{bmatrix}.
\end{align}

The resulting linear system is implemented using the AMDiS framework~\cite{AMDiS} and solved with the PETSc linear algebra backend~\cite{petsc-efficient, petsc-web-page}, using the MUMPS solver~\cite{MUMPS:1, MUMPS:2}.
\bigskip

\paragraph{Parallel implementation.} The formulation is well suited to distributed-memory parallelization by MPI. Under domain decomposition, the standard Stokes blocks $\vec{A}$ and $\vec{B}$ are assembled entirely from element-wise contributions and therefore require only the usual local finite-element assembly. The rigid-body coupling block $\vec{C}$ can likewise be assembled from local contributions on each subdomain, since its entries are of the form
\begin{align}
C_{ij} = -\int_{\Omega} \eta\, \boldsymbol{N}_j \cdot \vec{w}_i \ \mathrm{d}\Omega,
\end{align}
where \(\vec{w}_i\) denotes one of the rigid-body modes associated with translation or rotation. Hence $\vec{C}$ is obtained by standard parallel assembly of locally computed integrals.

The only genuinely global object is the small matrix $\vec{M}$, whose entries involve integrals over the whole domain, e.g.:
\begin{align}
M_{ij} = \int_{\Omega} \eta\, m_{ij}(\vec r)\, \ \mathrm{d}\Omega .
\end{align}
Its evaluation therefore requires a global reduction across MPI ranks. However, \(\vec{M}\) is of size \(3\times 3\) for a single squirmer in two dimensions (and remains block-structured and very small for multiple squirmers), so both its assembly and inversion are negligible compared to the solution of the Stokes system. In practice, the parallel cost is therefore dominated by the fluid solve, while the additional overhead associated with the rigid-body coupling is minimal.
\bigskip

\paragraph{Short-range hard-wall repulsive potential.}
Before advancing the squirmer center of mass $\vec x_c$, a short-range contact correction is applied to avoid overlap and to produce physically meaningful near-wall motion.
For each nearby obstacle, the closest point $\vec p$ to the squirmer center $\vec x_c$ is computed, with
\begin{align}
\vec g=\vec x_c-\vec p,\qquad
g=|\vec g|,\qquad
\hat{\vec{g}}=\frac{\vec g}{g},\qquad
h=g-a,
\end{align}
where $a$ is the squirmer radius and $h$ is the signed gap.

A short-range repulsive potential with prescribed cutoff $h_{\mathrm{cut}}$ is applied only when the squirmer is sufficiently close to the obstacle and approaching it, i.e. when $h < h_{\mathrm{cut}}$ and $\vec U\cdot\hat{\vec{g}} < 0$.

The center velocity $\vec{U}$ is decomposed into normal and tangential parts:
\begin{align}
\vec U_{\perp}=(\vec U\cdot\hat{\vec{g}})\,\hat{\vec{g}},
\qquad
\vec U_{\parallel}=\vec U-\vec U_{\perp}.
\end{align}
When the above cutoff condition is satisfied, the update suppresses/limits the inward normal contribution and keeps the tangential component, so that motion transitions to sliding along the obstacle instead of penetrating it. In practice, this yields: (i) no geometric overlap, (ii) robust behavior in narrow gaps, and (iii) near-wall kinematics dominated by tangential motion.
\bigskip 

\paragraph{Adaptive time-step update.}
After each solve, the time step is updated to keep the particle motion stable near obstacles, while avoiding unnecessary slowdown far from walls.  
The timestep $\Delta t$ is computed from two ingredients:
(i)~an interface-based bound proportional to $\xi / |\vec U\cdot \hat{\vec{g}}|$, and
(ii) a gap-based bound proportional to $g / |\vec U\cdot\hat{\vec{g}} |$. Let $\mathcal{N}$ be the number of neighboring obstacles to the particle and let  $\Delta t_{\xi,k}$ and $\Delta t_{g,k}$ be the timestep restrictions from (i) and (ii), respectively. 
The final step is the most restrictive admissible value over nearby obstacles, then clipped by global lower/upper safety bounds:
\begin{align}
\Delta t^{n+1}
=
\max\!\Bigl(\Delta t_{\min},
\min\!\bigl(\Delta t_{\max}^{\mathrm{free}},\,
\min_{k\in\mathcal N}\{\Delta t_{\xi,k},\Delta t_{g,k}\}\bigr)\Bigr),
\end{align}
with additional damping that limits abrupt changes relative to \(\Delta t^n\) (e.g. growth/shrink factors).  
This provides small steps during close approach and larger steps in free-swimming regimes.

\begin{figure}[t]
    \centering
    \begin{overpic}[width=\linewidth]{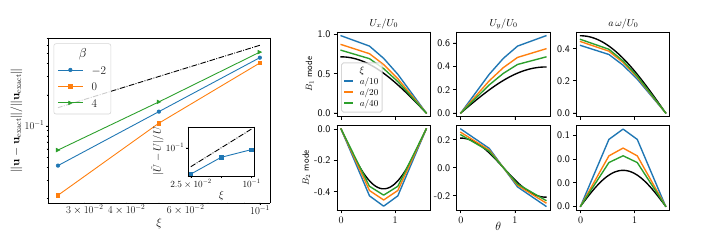}
    \put(0, 27){({\bf a})}
    \put(40, 27){({\bf b})}
    \end{overpic}
    \caption{({\bf a}): Relative error with respect to the analytical solution for selected values of $\beta$. The results show slightly better than first-order convergence, consistent with the phase-field approximation. (\emph{Inset}.) Relative error of the squirmer speed in free space compared with the exact value $U = B_1/2$. Linear convergence with respect to the interface thickness $\xi$ is observed, in agreement with the phase-field model.
    ({\bf b}) Translational and angular velocities of a squirmer near a wall for different interface thicknesses $\xi$. The top row shows $B_1 = 1$, $B_2 = 0$, while the bottom row shows $B_1 = 0$, $B_2 = 1$. Black lines denote the analytical solution~\cite{Ishimoto_Crowdy_2017}. The numerical results converge toward the analytical prediction as $\xi$ decreases.}    
    \label{fig:validation}
\end{figure}

\subsection{Validation}
\paragraph{Squirmer in free space.} As a first validation of the numerical method, we consider a single squirmer in free space. An analytical solution of the Stokes equations for a two-dimensional squirmer can be derived using a stream-function formulation in polar coordinates. The solution provides explicit expressions for the velocity field both inside and outside the squirmer disk. Typically, for squirmer motions only the first two modes are retained (i.e. $B_n = 0$ for $n>2$). In that case, the analytical solution inside the squirmer disk in the squirmer's reference frame is 
\begin{align*}
    u_{\rho}(\rho,\theta) &= \frac{B_1}{2}\left(1 - \left(\frac{\rho}{a}\right)^2\right)\cos\theta + B_2\frac{\rho}{a}\left(1 - \left(\frac{\rho}{a}\right)^2\right)\cos(2\theta), \\
    u_{\theta}(\rho,\theta) &= \frac{B_1}{2}\left(3\left(\frac{\rho}{a}\right)^2 - 1\right)\sin\theta + B_2\frac{\rho}{a}\left(2\left(\frac{\rho}{a}\right)^2 - 1\right)\sin(2\theta),
\end{align*}
whereas outside the squirmer disk the analytical solution is given by
\begin{align*}
    u_{\rho}(\rho,\theta) &= \frac{B_1}{2}\left(\left(\frac{a}{\rho}\right)^2 - 1\right)\cos\theta + B_2\frac{a}{\rho}\left(\left(\frac{a}{\rho}\right)^2 - 1\right)\cos(2\theta), \\
    u_{\theta}(\rho,\theta) &= \frac{B_1}{2}\left(\left(\frac{a}{\rho}\right)^2 + 1\right)\sin\theta + B_2\left(\frac{a}{\rho}\right)^3\sin(2\theta).
\end{align*}

Direct comparison with the analytical solution in the unbounded domain is not appropriate for numerical simulations, which are necessarily performed in bounded computational domains. Boundary conditions imposed at the outer boundary introduce deviations from the infinite-domain solution. For this reason, we validate the method by comparing the numerical velocity field with the analytical solution inside the squirmer disk, where both descriptions are consistent. Therefore, we define the relative error as
\begin{align}
E_{\mathrm{rel}} =
\frac{\|\vec{u}-\vec{u}_{\mathrm{exact}}\|_{L^2_{\phi}(\Omega)}}
{\|\vec{u}_{\mathrm{exact}}\|_{L^2_{\phi}(\Omega)}},
\qquad
\|f\|_{L^2_{\phi}(\Omega)} =
\left(\int_{\Omega} \phi(\vec{x}) |f(\vec{x})|^2\,\mathrm{d}\Omega\right)^{1/2},
\end{align}
where $\vec{u}$ denotes the numerical velocity field and $\vec{u}_{\mathrm{exact}}$ the analytical solution. Further, we compare the free space speed of the squirmer $U = B_1/2$ with the one computed numerically. The results are shown in Fig.~\ref{fig:validation}~{\bf a}, showing nice convergence and relative errors of order $\mathcal{O}(10^{-2})$ for small interface thickness $\xi$. 
\bigskip

\paragraph{Squirmer near a wall.} We further validate our model by computing the instantaneous translational and rotational velocities of a squirmer near a no-slip wall and comparing them to the analytical results presented in~\cite{Ishimoto_Crowdy_2017}:
\begin{subequations}
\begin{align}
U_x &= \frac{1-\rho^2}{2}\left[B_1 \cos\theta -2\rho B_2\sin (2\theta)\right]\\
U_y &= \frac{(1 - \rho^2)^2}{2(1 + \rho^2)} \left[B_1\sin\theta + 2\rho B_2\cos(2\theta)\right]\\ 
\omega &= \frac{\rho^2}{a(1+\rho^2)}\left[2\rho B_1 \cos\theta + (1 - 3\rho^2) B_2\sin(2\theta)\right],
\end{align}
\end{subequations}
where $\rho = d/a - (d^2/a^2 - 1)^{1/2}$, and $d$ is the distance of the center of mass $\vec{x}_C$ to the wall. 

We fix the distance of the squirmer' s center of mass to $d = 1.2 a$ from the wall and vary its orientations as 
\[
\theta \in\{0, \pi/6, \pi/4, \pi/3,\pi/2 \}.
\] 
For each orientation, we compute the translational and angular velocities for two sets of  squirmer modes $(B_1 = 1, B_2 = 0)$ and $(B_1 = 0, B_2=1)$. These computations are repeated for decreasing values of the interface thickness $\xi = a/10$, $a/20$ and $a/40$. The results are shown in Fig.~\ref{fig:validation}~{\bf b}, where we observe that decreasing $\xi$ leads to numerical solutions that closely match the analytical predictions.

\paragraph{Squirmer interacting with one obstacle}

\begin{figure}[ht]
    \centering
    \includegraphics[width=0.75\linewidth]{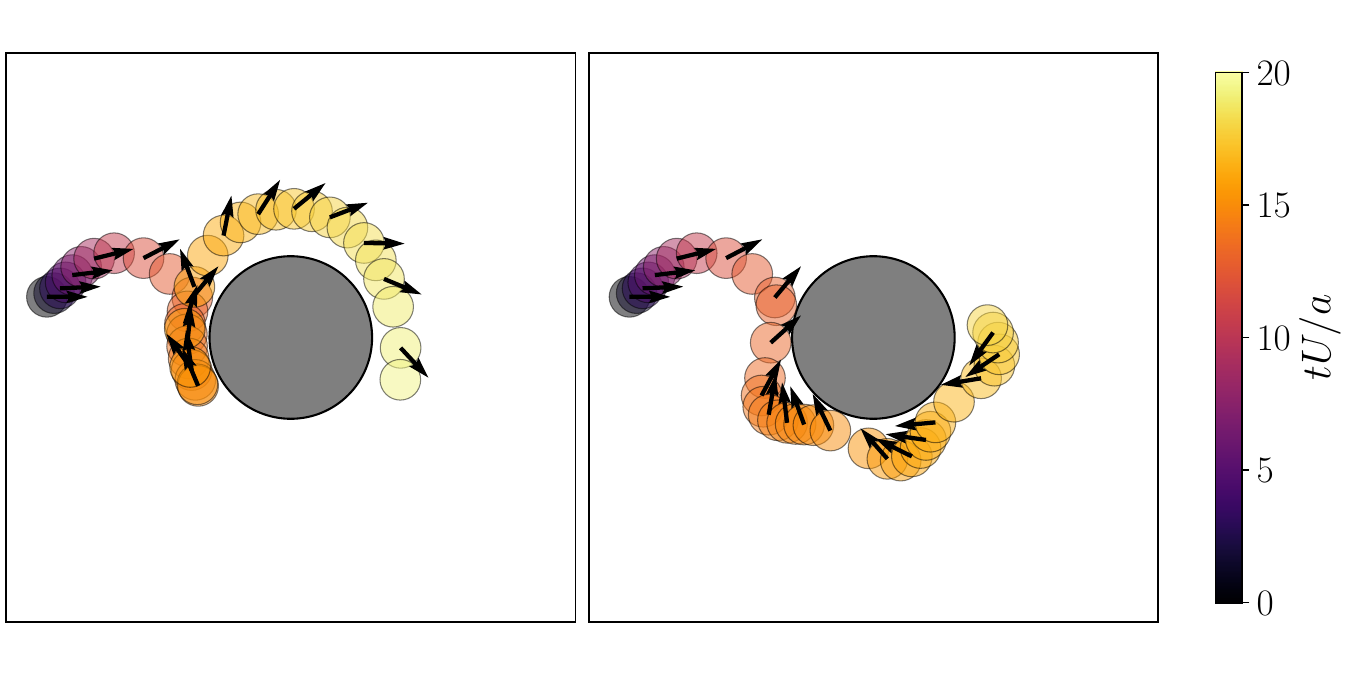}
    \caption{
Interaction of a pusher ($\beta=-10$) with a circular obstacle of radius $R=4a$.
\emph{Left:} Forward orbit obtained for a repulsive cutoff gap $h_{\mathrm{cut}}=a/4$.
\emph{Right:} Backward orbit obtained for $h_{\mathrm{cut}}=a/20$. The squirmer is color coded according to the rescaled time $t U/a$, with $a$ the radius of the squirmer and $U=B_1/2$ the free-space velocity. The black arrows indicate the squirmer orientation. 
Qualitatively similar behaviors were reported in Ref.~\cite{Kuron_2019}.
}
    \label{fig:one_obstacle}
\end{figure}

We investigate the interaction between a squirmer of radius $a$ and a circular obstacle of radius $R=4a$. Simulations are performed in a square domain of size $L=64a$, with no-slip boundary conditions imposed both on the obstacle surface and on the domain boundaries. To prevent overlap between the squirmer and the obstacle, we employ a short-range hard-wall repulsive potential characterized by a cutoff gap $h_{\mathrm{cut}}$. Two values are considered: $h_{\mathrm{cut}}=a/20$ and $h_{\mathrm{cut}}=a/4$.

Consistent with the results reported in~\cite{Kuron_2019}, we observe two distinct dynamical regimes. For small values of $|\beta|$, the squirmer scatters from the obstacle, whereas for sufficiently large $|\beta|$ it becomes hydrodynamically trapped and orbits around it (results not shown). In the orbiting regime, the dynamics depend on the swimmer type. Pullers ($\beta>0$) exhibit forward orbits, with their orientation aligned with the direction of motion, and display periodic oscillations whose amplitude gradually decays. Pushers ($\beta<0$), in contrast, show a behavior that depends on the cutoff gap $h_{\mathrm{cut}}$. For the smaller gap $h_{\mathrm{cut}}=a/20$, the swimmer tends to orient opposite to its direction of motion, resulting in a backward orbit. For the larger gap $h_{\mathrm{cut}}=a/4$, the motion resembles that of pullers: after a short transient backward motion, the swimmer settles into a forward orbit. An example of these behaviors is shown in Fig.~\ref{fig:one_obstacle}.

\subsection{Classification of particle trajectories}

Particle trajectories are classified according to the long-time behavior of the displacements. Let $t_f$ be the final time of the trajectory. We consider the squared displacement measured relative to the final position, $
d^2(t) = |\vec x_C(t)-\vec x_C(t_f)|^2$.

A trajectory is considered \textit{confined} if this squared displacement remains bounded over a sufficiently long time interval preceding $t_f$. To ensure that this bounded behavior is not transient, the analysis is applied iteratively to progressively shorter segments of the trajectory approaching the final time.

Trajectories that do not exhibit bounded displacement over these time intervals are classified as \textit{free}, corresponding to particles whose displacement continues to grow over the observation time.

For confined trajectories we further distinguish two regimes according to the spatial extent of the motion. Let $d_{\max}$ denote the characteristic amplitude of the bounded displacement. If $d_{\max}$ is smaller than $a/10$, where $a$ is the particle radius, the particle effectively remains localized and the trajectory is classified as \textit{static trapping}. Larger values correspond to \textit{dynamical trapping}, where the particle remains confined but continues to move within a finite region.
\end{document}